\documentclass[apjl]{emulateapj}
\usepackage{apjfonts}

\newcommand{\hii}{H\,{\sc ii}\rm}

\slugcomment{}

\shorttitle{The effect of blending on the
Cepheid distance to NGC~300} \shortauthors{Bresolin, Pietrzy\'nski,
Gieren \& Kudritzki}

\begin{document}


\title{The Araucaria Project: The effect of blending on the Cepheid
distance to NGC~300 from Advanced Camera for Surveys
images$^1$}\footnotetext[1]{Based on observations with the NASA/ESA
{\em Hubble Space Telescope}, obtained at the Space Telescope Science
Institute, which is operated by AURA, Inc., under NASA contract
NAS5-26555.  These observations are associated with program 9492.}

\author{Fabio Bresolin} \affil{Institute for Astronomy, 2680 Woodlawn
Drive, Honolulu, HI 96822; bresolin@ifa.hawaii.edu}

\author{Grzegorz Pietrzy\'nski} \affil{Departamento de Fisica, Astronomy
Group, Universidad de Concepci\'on, Casilla 160-C, Concepci\'on,
Chile}

\affil{Warsaw University Observatory, Al. Ujazdowskie 4,PL-00-478,
Warsaw, Poland; pietrzyn@hubble.cfm.udec.cl}

\author{Wolfgang Gieren} \affil{Departamento de Fisica, Astronomy
Group, Universidad de Concepci\'on, Casilla 160-C, Concepci\'on,
Chile; wgieren@astro-udec.cl}

\and 

\author{Rolf-Peter Kudritzki} \affil{Institute for Astronomy, 2680 Woodlawn
Drive, Honolulu, HI 96822; kud@ifa.hawaii.edu}

\begin{abstract} 

We have used the Advanced Camera for Surveys aboard the Hubble Space
Telescope to obtain $F435W$, $F555W$ and $F814W$ single-epoch images
of six fields in the spiral galaxy NGC~300.  Taking advantage of the
superb spatial resolution of these images, we have tested the effect
that blending of the Cepheid variables studied from the ground with
close stellar neighbors, unresolved on the ground-based images, has on
the distance determination to NGC~300. Out of the 16 Cepheids included
in this study, only three are significantly affected by nearby stellar
objects. After correcting the ground-based magnitudes for the
contribution by these projected companions to the observed flux, we
find that the corresponding Period-Luminosity relations in $V$, $I$
and the Wesenheit magnitude $W_I$ are not significantly different from
the relations obtained without corrections. We fix an upper limit of
0.04 magnitudes to the systematic effect of blending on the distance
modulus to NGC~300.

As part of our HST imaging program, we present improved photometry for
40 blue supergiants in NGC~300.

\end{abstract}



\section{Introduction}

Cepheid variables continue to play an important role in defining the
extragalactic distance scale, as demonstrated by the improvements
obtained in the past few years in the calibration of far-reaching
secondary distance indicators by the HST Key Project
(\citealt{freedman01}) and the growing samples of variables with
high-quality ground-based light curves located in galaxies of the
Local Group and out to distances of a few Mpc (\citealt{udalski99},
\citealt{pietrzynski02, pietrzynski04}).  Continuing efforts are spent
to solve some of the pending issues related to sources of potential
systematic errors in the distances derived from the Cepheid
Period-Luminosity (PL) relation. Among the recent work, we recall the
confirmation, based on interferometric data (\citealt{nordgren02}), of
the \citet{fouque97} calibration of the infrared surface brigthness
technique, which allows to derive accurate geometric distances to
Cepheids. This result, together with the availability of excellent
near-infrared photometry of Cepheids in the LMC, allowed
\citet{gieren05a} to exclude a suspected dependence of the slope of
the Cepheid PL relation on the metallicity of the host galaxy.

Systematic uncertainties can affect the extragalactic distances based
on Cepheids in additional, subtle ways. The problem of the metallicity
dependence of the zero point of the PL relation, for example, has
eluded a satisfying solution for a long time, generating controversies
on the magnitude and direction of the effect. Among the most recent
reassessments of the problem is the work by \citet{sakai04}, who found
that a change in metal abundance by one order of magnitude has an
effect on the distance modulus derived from PL relations in the
optical $V$ and $I$ passbands of $-0.24$ magnitudes.

Blending of Cepheids with unresolved projected companions is also a
source of concern, because our increasing inability to spatially
resolve and account for such companions with increasing distances
could lead to systematically underestimate distances to galaxies.
\citet{stanek99}, followed by \citet{mochejska00,mochejska01} as part
of the DIRECT project on M31 and M33, drew the attention to the
effects of blending on the distances to the HST Key Project galaxies,
which have been derived neglecting this effect. By extrapolating the
results obtained in the Local Group galaxies, these authors suggested
an upward correction to the distance moduli, exceeding 0.3 magnitudes
at 25 Mpc, implying that the Hubble constant could be overestimated by
5\%--10\%. These results have been questioned by \citet{gibson00} and
\citet{ferrarese00}, on the basis of artificial star experiments and
looking for distance-dependent residuals in the Tully-Fisher relation
for the Key Project galaxies.  A maximum bias due to blending of $\sim
0.02$ magnitudes was found.

In the Araucaria
Project\footnote{http://ifa.hawaii.edu/$\sim$bresolin/Araucaria/}
(\citealt{gieren01}) we are endeavoring to obtain accurate Cepheid
distances to galaxies in the Local Group and in the Sculptor Group,
and to decrease the uncertainties associated with some of the most
important surces of systematic errors. Stellar spectroscopy in the
target galaxies provides us with metal abundances of blue supergiants,
which complement \hii\/ region oxygen abundances, and which are
necessary in order to constrain the metallicity dependence of the PL
relation.  One of the key target galaxies in the Araucaria Project is
NGC~300 in Sculptor. For this galaxy we have an excellent
characterization of both the Cepheid (\citealt{pietrzynski02}) and the
blue supergiant (\citealt{bresolin02}) populations. Follow-up work
allowed us to measure an accurate distance from optical and
near-infrared photometry (\citealt{gieren04,gieren05}) and to derive
the stellar metal abundance gradient of NGC~300
(\citealt{urbaneja05}). While we had reasons to believe that blending
of Cepheids with nearby stars has a negligible impact on our distance
to NGC~300 (\citealt[hereafter G04]{gieren04}), a direct test made on
high-resolution HST images was still desirable.  We therefore obtained
data with the Advanced Camera for Surveys (ACS), with the goal of
improving the ground-based photometry of a significant sample of
Cepheids (for the blending problem) and blue supergiants (for the use
of these stars as independent stellar distance indicators,
\citealt{kudritzki03}) in NGC~300.

At a distance of $\sim2$ Mpc, one ACS pixel (0\farcs05) corresponds to
$\sim0.5$ pc. The HST images discussed in this paper therefore allow
us to follow a straightforward empirical approach to test in detail
for the presence of projected luminous stars that could affect the
photometry of the Cepheid variables included in the PL relation.  The
paper is organized as follows: we discuss the ACS data and the stellar
photometry in \S2, and the photometry of the Cepheids in particular,
together with the comparison with the ground-based light curves, in
\S3. We describe the method adopted to quantify the effects of
blending on the Cepheid photometry in \S4. The impact of blending on
the Cepheid PL relation is presented in \S5. We briefly discuss and
summarize our results in \S6. We include in the Appendix a table with
the HST photometry of 40 blue supergiants.


\section{Data}

\begin{figure*}
\epsscale{0.9}
\plotone{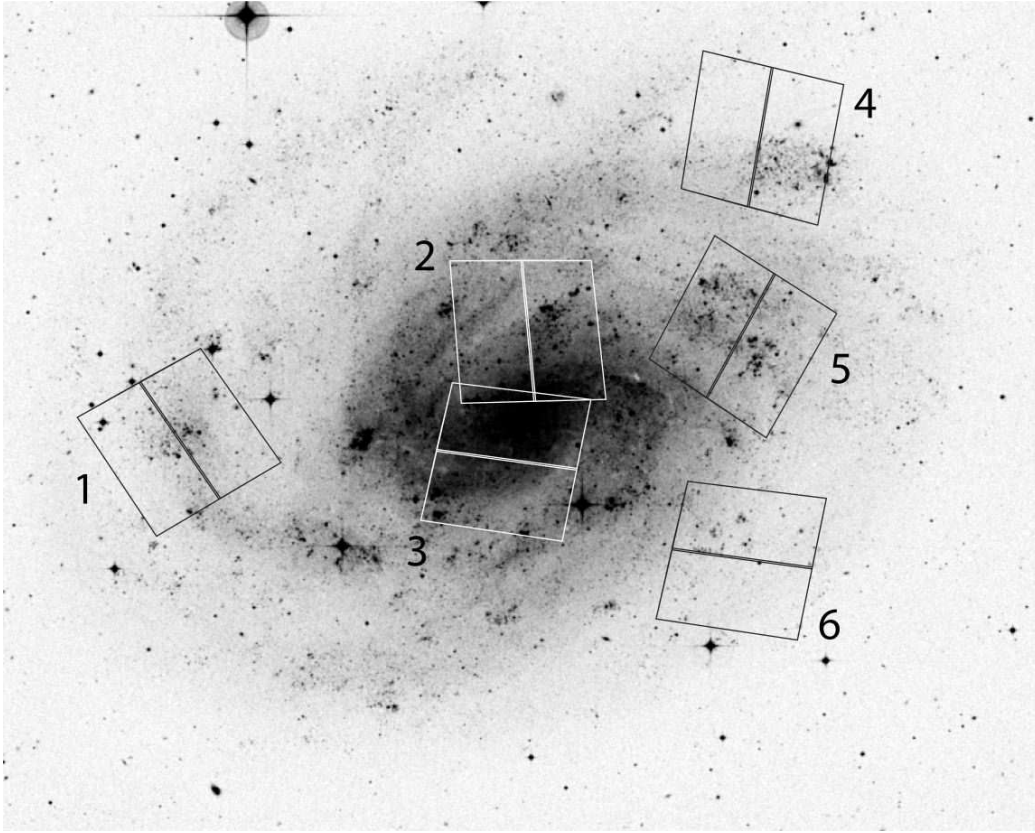}
\vspace{7mm}
\caption{The footprints of the six ACS fields overlaid on a 
Digital Sky Survey image of NGC~300.}
\label{map}
\epsscale{1}
\end{figure*}

Observations of six ACS fields in NGC~300 were obtained during HST
Cycle 11 as part of program GO-9492 (PI: Bresolin) from July 2002 to
December 2002 (see Table~\ref{log} for a journal of the
observations). Wide Field Camera images, approximately
202\arcsec$\times$202\arcsec\/ on the sky, were obtained through the
$F435W$, $F555W$ and $F814W$ filters. The exposure time was 1080
seconds in the $F435W$ and $F555W$ filters, subdivided into three
dithered exposures. Four exposures, totalling 1440 seconds, were
obtained through $F814W$. These exposures were accommodated in
two-orbit HST visits for each field. The goal of the program was to
obtain accurate photometry of a significant fraction of the blue
supergiants and Cepheid variables studied in our previous
spectroscopic and photometric surveys in NGC~300
(\citealt{bresolin02}, \citealt{pietrzynski02}).  The fields were
therefore selected in order to sample the population of supergiants
and Cepheids at different galactocentric distances. Fig.~\ref{map}
shows the ACS footprints of the six fields on a DSS image of the
galaxy.

\begin{deluxetable}{clc}
\tablewidth{0pt}
\tablecolumns{3}
\tablecaption{Observation journal\label{log}}
\tablehead{
\colhead{ACS Field no.}            &
\colhead{UT date}    &
\colhead{JD\tablenotemark{a}}    }
\startdata
\vspace{-2mm}
\\
1\dotfill  &  2002 Jul 17  &  2,452,472.85 \\
2\dotfill  &  2002 Jul 19  &  2,452,474.78 \\
3\dotfill  &  2002 Sep 28  &  2,452,545.63 \\
4\dotfill  &  2002 Jul 21  &  2,452,476.79 \\
5\dotfill  &  2002 Dec 25  &  2,452,633.12 \\
6\dotfill  &  2002 Sep 26  &  2,452,543.56 \\
\enddata
\tablenotetext{a}{Mean Julian Date for the two-orbit HST visits}
\end{deluxetable}

Stellar photometry was carried out on the pipeline-processed images
using the DOLPHOT\footnote{http://purcell.as.arizona.edu/dolphot/.}
(version 1.0) package, an adaptation of HSTphot (\citealt{dolphin00})
to ACS images. For our final photometry, obtained with the available
pre-computed Point Spread Functions, we retained all objects that were
identified as well-fitted stars in all three filters.  In order to allow a
meaningful comparison with our previous ground-based photometry, we
have transformed the magnitudes from the HST system to \mbox{\em BVI},
following the transformation equations published by
\citet{sirianni05}, although this procedure necessarily introduces
additional uncertainty in the photometric measurements.

\begin{figure}[t]
\plotone{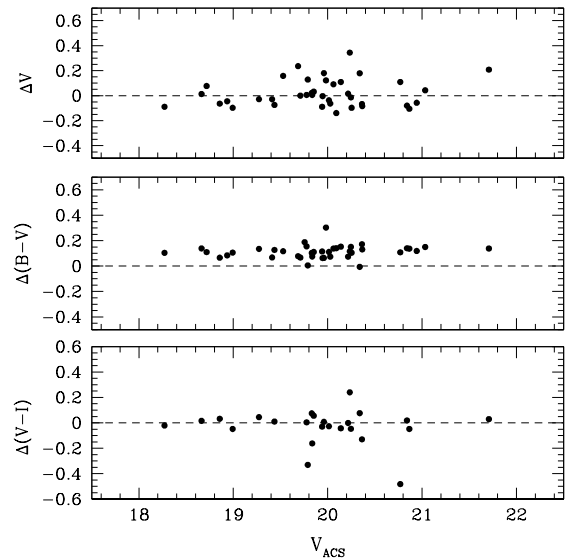}
\caption{Comparison between the ground-based photometry of
\citet{bresolin02} and the ACS photometry of the 40 blue supergiants in common.
Along the vertical axis we show 
the difference (ACS $-$ ground-based) of the following quantities: $V$ ({\em
top}), $B-V$ ({\em middle}) and $V-I$ ({\em bottom}). Only 24 stars
have $I$ magnitudes available from the ground.
}
\label{comp}
\end{figure}

A comparison between the ACS and the ground-based $V$ magnitudes for
the 40 blue supergiants in common with \citet{bresolin02} reveals no
significant zero-point offset (Fig.~\ref{comp}, top). A few stars
appear slightly fainter in the ACS photometry, as expected from the
higher spatial resolution attained from HST, which resolves some of
the supergiants from projected close companions, however the level of
contamination is small for supergiant stars ($V<21$). We note that at
least part of the scatter in this plot is likely due to the stochastic
variability (at the 0.1 mag level) detected for the blue supergiant
stars in NGC~300 by \citet{bresolin04}. The fact that the color
difference plots (lower two panels in Fig.~\ref{comp}) show a smaller
scatter supports this conclusion (the variability is similar in the
different passbands).  No magnitude offset is seen in the $V-I$ color
comparison (bottom panel of Fig.~\ref{comp}; 24 stars are shown, due
to the smaller number of stars with ground-based $I$ magnitudes
relative to $B$ and $V$). A significant ($\sim0.1$ mag) zero-point
difference is present in $B-V$ (middle panel). This discrepancy
appears to be due to a problematic calibration of the $B$ filter of
our ground-based photometry, which was already noted by
\citet{bresolin04}. The main outliers in the $V-I$ color comparison
are due to extreme red colors in the ground-based photometry (stars
C-5 and C-11), whereas the HST photometry provides more consistent
colors for their early-B spectral types and their $B-V$.

A table summarizing the ACS photometry of the blue supergiants is
presented in the Appendix.

\begin{deluxetable}{ccccc}
\tablewidth{0pt}
\tablecolumns{5}
\tablecaption{ACS photometry of NGC~300 Cepheids\label{magnitudes}}
\tablehead{
\colhead{ID}            &
\colhead{ACS field}	&
\colhead{$B$}    &
\colhead{$V$}    &
\colhead{$I$}}
\startdata
\vspace{-2mm}
\\
cep004 & 2 & 21.07 & 19.97 & 18.98 \\
cep007 & 5 & 21.26 & 20.48 & 19.71 \\
cep015 & 2 & 22.71 & 21.56 & 20.50 \\
cep018 & 4 & 22.19 & 21.33 & 20.47 \\
cep022 & 1 & 22.58 & 21.62 & 20.65 \\
cep028 & 2 & 22.23 & 21.40 & 20.56 \\
cep032 & 2 & 22.64 & 21.56 & 20.62 \\
cep035 & 2 & 21.85 & 21.23 & 20.61 \\
cep040 & 5 & 23.14 & 22.24 & 21.29 \\
cep044 & 3 & 22.32 & 21.58 & 20.81 \\
cep045 & 1 & 22.90 & 22.11 & 21.24 \\
cep050 & 6 & 23.56 & 22.44 & 21.53 \\
cep051 & 4 & 23.25 & 22.20 & 21.22 \\
cep066 & 1 & 22.28 & 21.76 & 21.09 \\
cep067 & 5 & 23.20 & 22.31 & 21.39 \\
cep074 & 4 & 22.80 & 22.09 & 21.35 \\
\enddata
\end{deluxetable}

\section{Cepheid photometry} 

Single-epoch photometry is available from our ACS observations for 30
of the NGC~300 Cepheids included in the catalog of
\citet{pietrzynski02}. The latter work provided high-quality $B$ and
$V$ light curves for 117 Cepheids, observed at the 2.2\,m ESO/MPI
telescope on La Silla. We adopt here the nomenclature introduced in
that work to identify the Cepheids.  Additional $V$ and $I$ data were
obtained by G04 at Las Campanas and Cerro Tololo, in order to improve
the $V$ light curves and to obtain $I$ data to address the reddening
correction issue.  For the rest of this paper we will limit our
analysis to 16 variables in common with the work of G04, which
includes only Cepheids with periods longer than 10 days, and which
were used for the derivation of the distance to NGC~300 based on the
Period-Luminosity relation in $V$ and $I$.

\begin{figure} \plotone{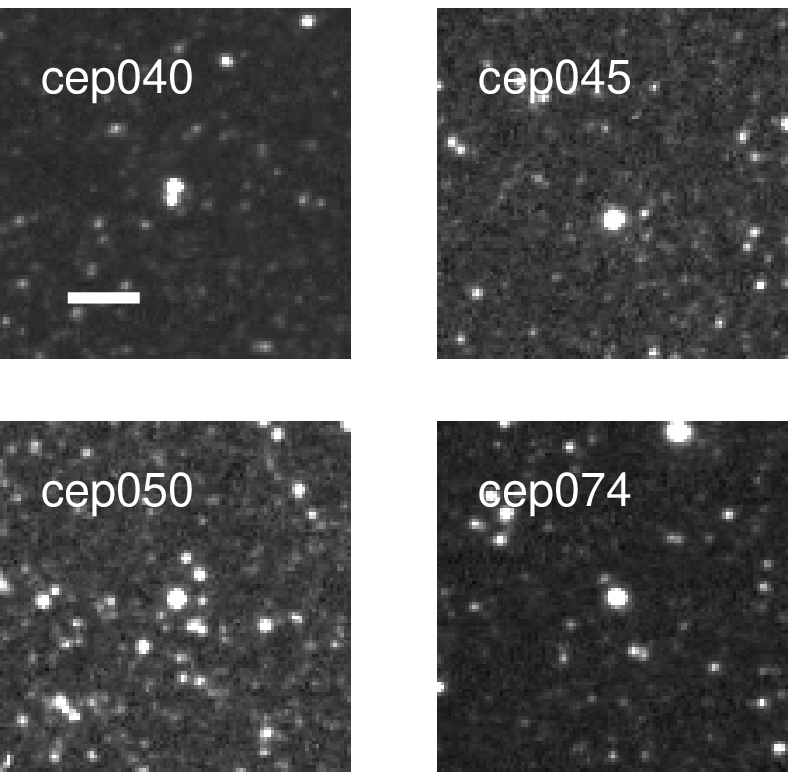} \caption{Close companions to
Cepheid variables are uncovered from the ACS images, as shown in these
5\arcsec$\times$5\arcsec\/ frames, centered on the stars cep040,
cep045, cep050 and cep074, and produced by combining the available
$F555W$ images with the PyDrizzle software. The horizontal bar is
1\arcsec\/ long. Contrary to the remaining cases, the faint
companions to cep074 do not contribute significantly to the total flux
of the blend.  \label{4cepheids}} \end{figure}

The {\em BVI} ACS magnitudes of the Cepheids are presented in
Table~\ref{magnitudes}. These magnitudes are compared with the
ground-based light curves in Fig.~\ref{ceph}, where the phases of the
HST data have been calculated from the periods and ephemerides
published by \citet{pietrzynski02}. We updated the periods of cep004
(from 75.0 to 79.5 days) and cep007 (from 43.35 to 43.20), which
improved the fit to the ground-based observations. As Fig.~\ref{ceph}
shows, we find excellent agreement between the ACS data points and the
ground-based data in $V$ and $I$. The $B$ data indicate the presence
of the same offset noticed in the blue supergiant magnitude comparison
done in the previous section.  Some notable exceptions to the good
agreement can also be found, for example cep040 and cep050, for which
the ACS magnitudes are systematically fainter in all bands by several
tenths of a magnitude. This is the effect we expect from the presence
of unresolved blends in the ground-based data. Close companions to the
stars mentioned above are indeed detected on the ACS images, as shown
in Fig.~\ref{4cepheids}.

\begin{figure*}
\epsscale{0.9}
\plotone{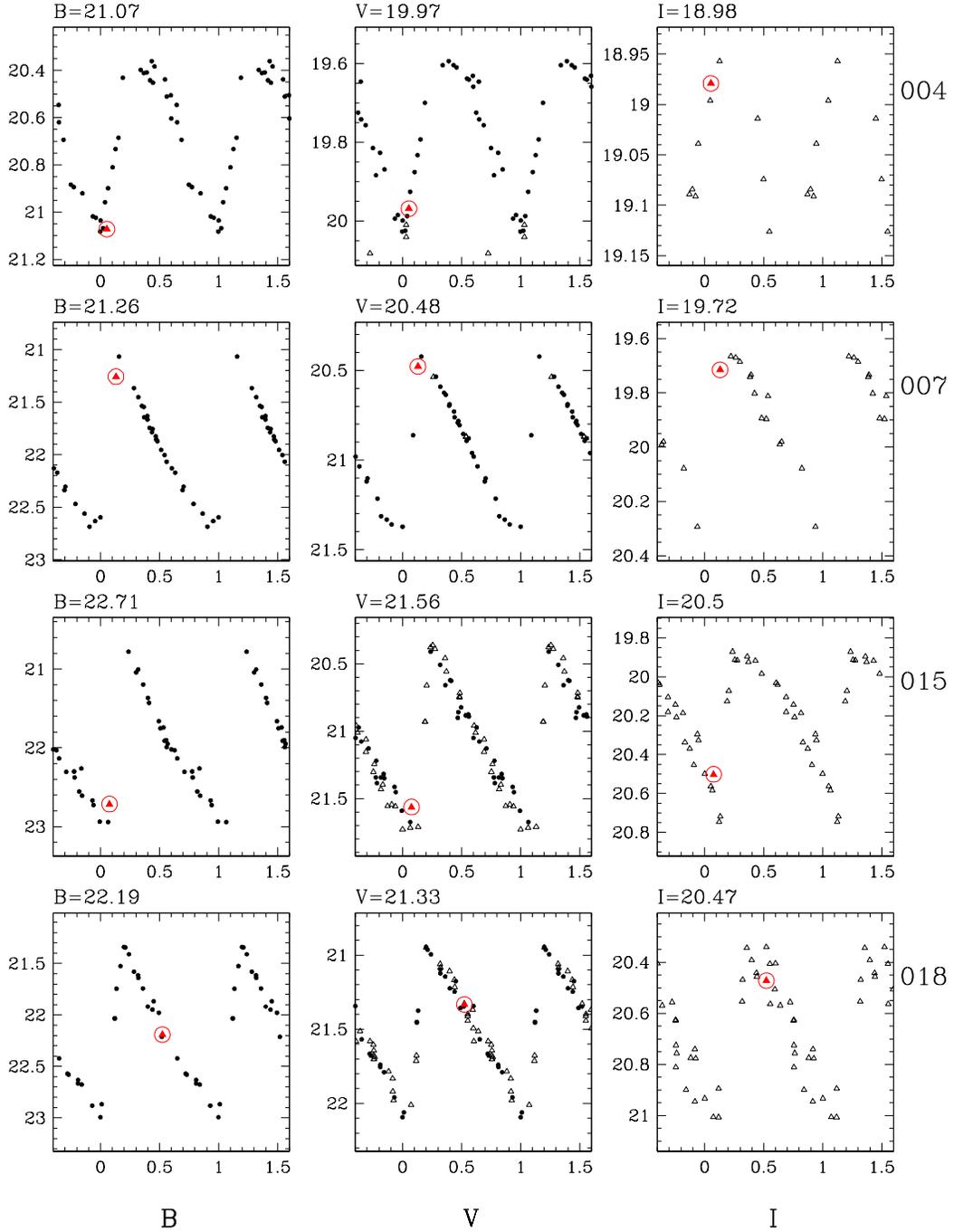}
\caption{The ACS single-epoch photometry (triangle-in-circle symbol) 
plotted on the
ground-based light curves for the NGC~300 Cepheids in common. The
ground-based photometry is from \citet{pietrzynski02} (full dots),
complemented by additional $VI$ data from
G04 (open triangles). Each row shows, for a given
Cepheid, the data relative to the $B$, $V$ and $I$ magnitudes. The ACS
magnitudes for each variable are shown above each plot.
The Cepheid identification
scheme follows \citet{pietrzynski02}, and is indicated on the right of
each row. 
}
\label{ceph}
\end{figure*}

\begin{figure*}
\epsscale{0.9}
\figurenum{3}
\plotone{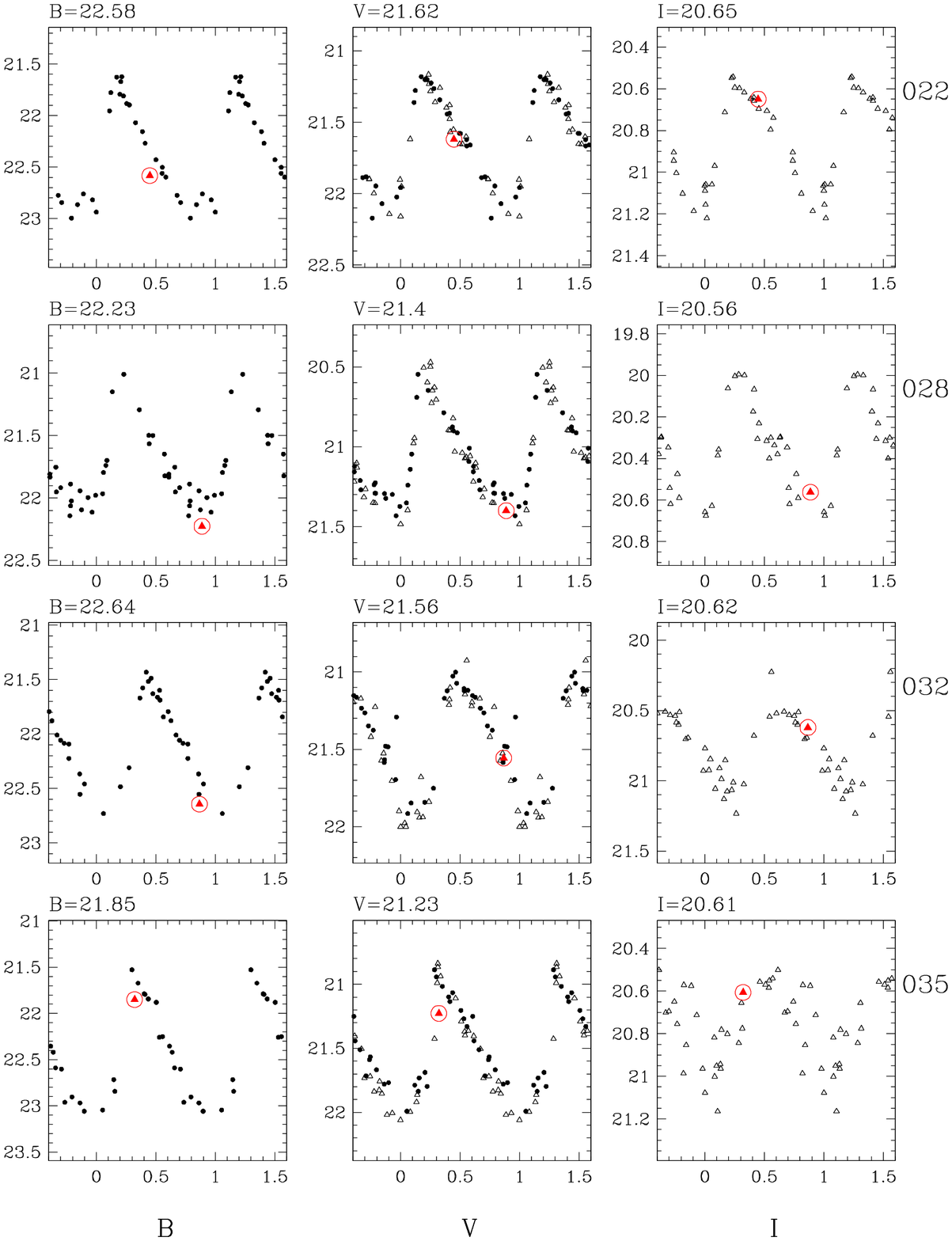}
\caption{{\em Continued}
}
\end{figure*}

\begin{figure*}
\epsscale{0.9}
\figurenum{3}
\plotone{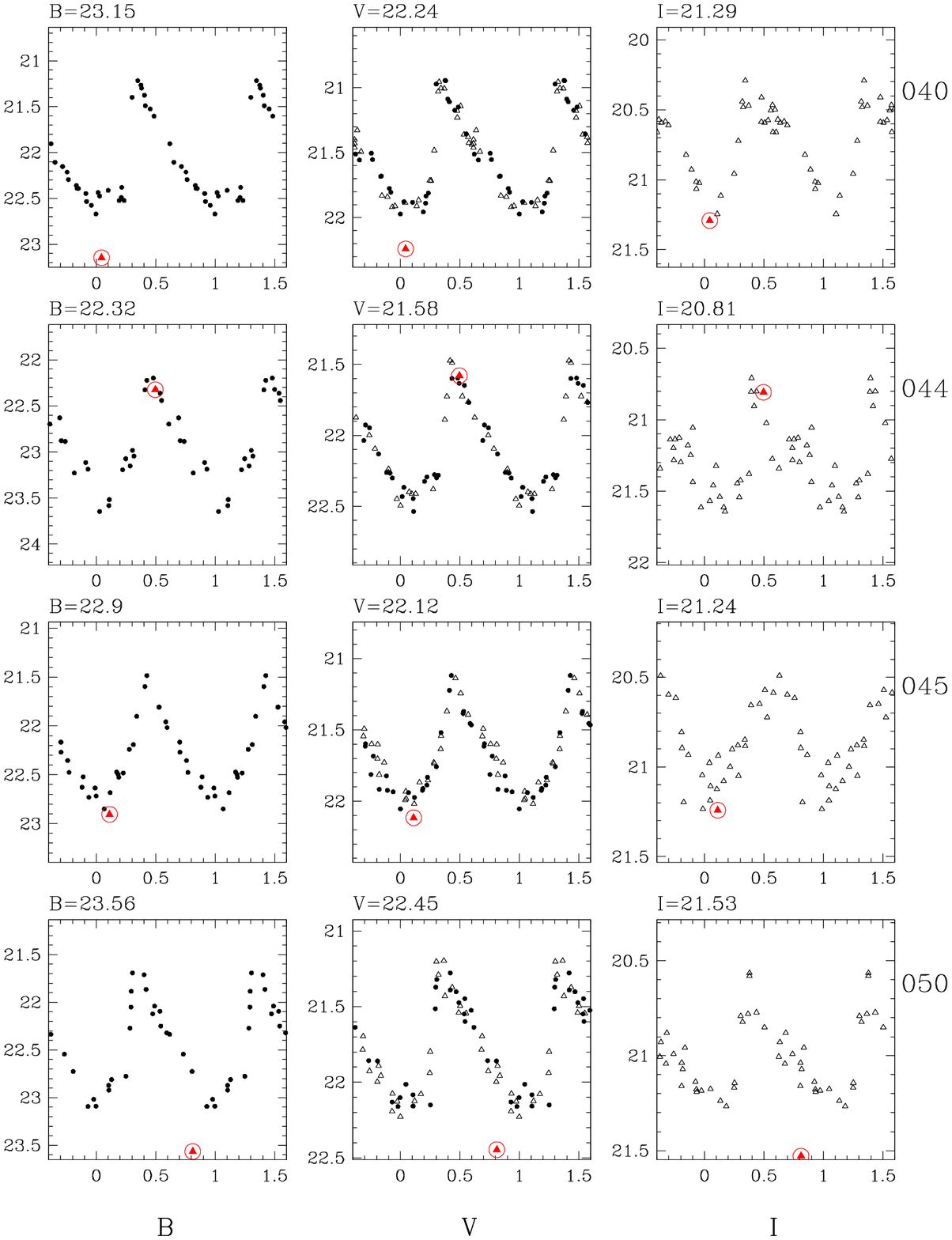}
\caption{{\em Continued}
}
\end{figure*}

\begin{figure*}
\epsscale{0.9}
\figurenum{3}
\plotone{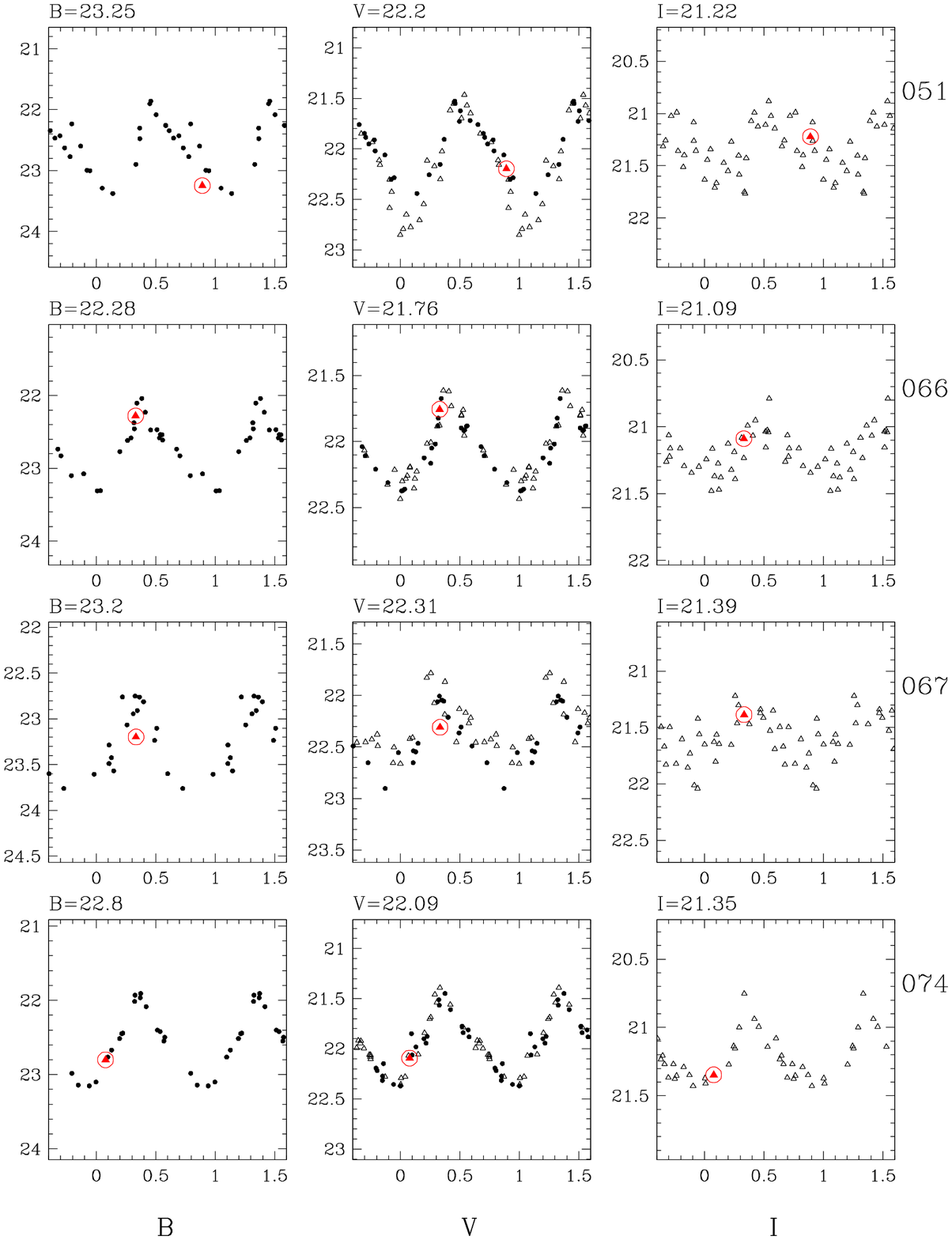}
\caption{{\em Continued}
}
\end{figure*}
\epsscale{1.0}

\section{Definition and characterization of blends}

In order to study the effects of blending on the ground-based
P-L relation, we have to consider the contribution to the observed
Cepheid flux by unresolved projected companions. The ACS images,
thanks to their superior spatial resolution, allow us to estimate this
contribution, under the assumption that the variables are not further
blended in the HST images.

We have adopted a straightforward empirical method to estimate the
amount of photometric contamination for the Cepheids. For each
variable, a Fourier series analysis has provided a fit to the $V$ and
$I$ light curves. We have then determined the offset between the ACS
magnitude and that provided by the Fourier fit, calculated at the
corresponding phase value. Whenever this difference was smaller than
the 1$\sigma$ dispersion of the Cepheid data points around the fit (in
the range 0.06--0.13 mag), we assigned a zero value for the
photometric contamination by blends. Larger offsets resulted in our
estimate for the non-zero effect of nearby companions on the
ground-based photometry.

In an alternative approach, we could have measured the total flux
contributed by companions, resolved by HST, within a pre-specified
distance from the Cepheid. This is the method that was adopted by
\citet{mochejska00,mochejska01} as part of the DIRECT Project study of
the effects of blending in M31 and M33. However, the maximum distance
from the Cepheid that a star needs to have in order to be considered
blended with it, around half the typical FWHM of the Point Spread
Function (PSF) of the stellar images in the ground-based images, is a
somewhat subjective parameter, and can vary between images taken under
different observing conditions. It is also necessary to impose a lower
limit to the flux of the stars included in the count, because faint
stars resolved by HST would not be measurable from the ground, and
would instead contribute to the galaxy background flux. Without an
{\em a priori} knowledge of all these quantities it would be difficult
to obtain a correct estimate of blending.  This is especially true in
our case, where observations were diluted over a large temporal
interval and were obtained with a variety of telescopes and
cameras. We have chosen to estimate these parameters {\em a
posteriori}, once the analysis based on the displacement from the mean
light curves had been carried out. We have used the HST photometry of
stars around the position of the Cepheids to check our results, to
verify that no bright companion was unaccounted for, and that we could
identify the companions associated with the offsets between the ACS
and the ground-based photometry.  This procedure also allowed us to
resolve a few dubious cases, where the Fourier analysis fits provided
uncertain results, due for example to a scarcely populated steep
ascending branch of the Cepheid light curve.

We found that only three out of 16 Cepheids required a significant
(i.e.~larger than the uncertainties) magnitude correction in order to
bring the HST and the ground-based photometry into agreement. Although
this small fraction might appear surprising at first, we should recall
that the Cepheids used for the determination of the distance to
NGC~300 by G04 (the source of the ground-based data used in this work)
are among those with the best available characteristics, such as
well-defined and clean light curves, and that for the PL relation the
most discrepant objects, which are potential blends, are rejected.
Cepheids having a mean magnitude brighter than $V=22$ are not affected
by blends, i.e.~we find excellent agreement between the HST photometry
and the ground-based photometry. The affected objects (cep040, cep045
and cep050) are among the faintest in our sample, spanning the range
of mean magnitudes between $V=22.1$ and $V=22.4$. The corresponding
magnitude offsets are summarized in Table~\ref{blends}.

\begin{figure} 
\bigskip
\bigskip
\plotone{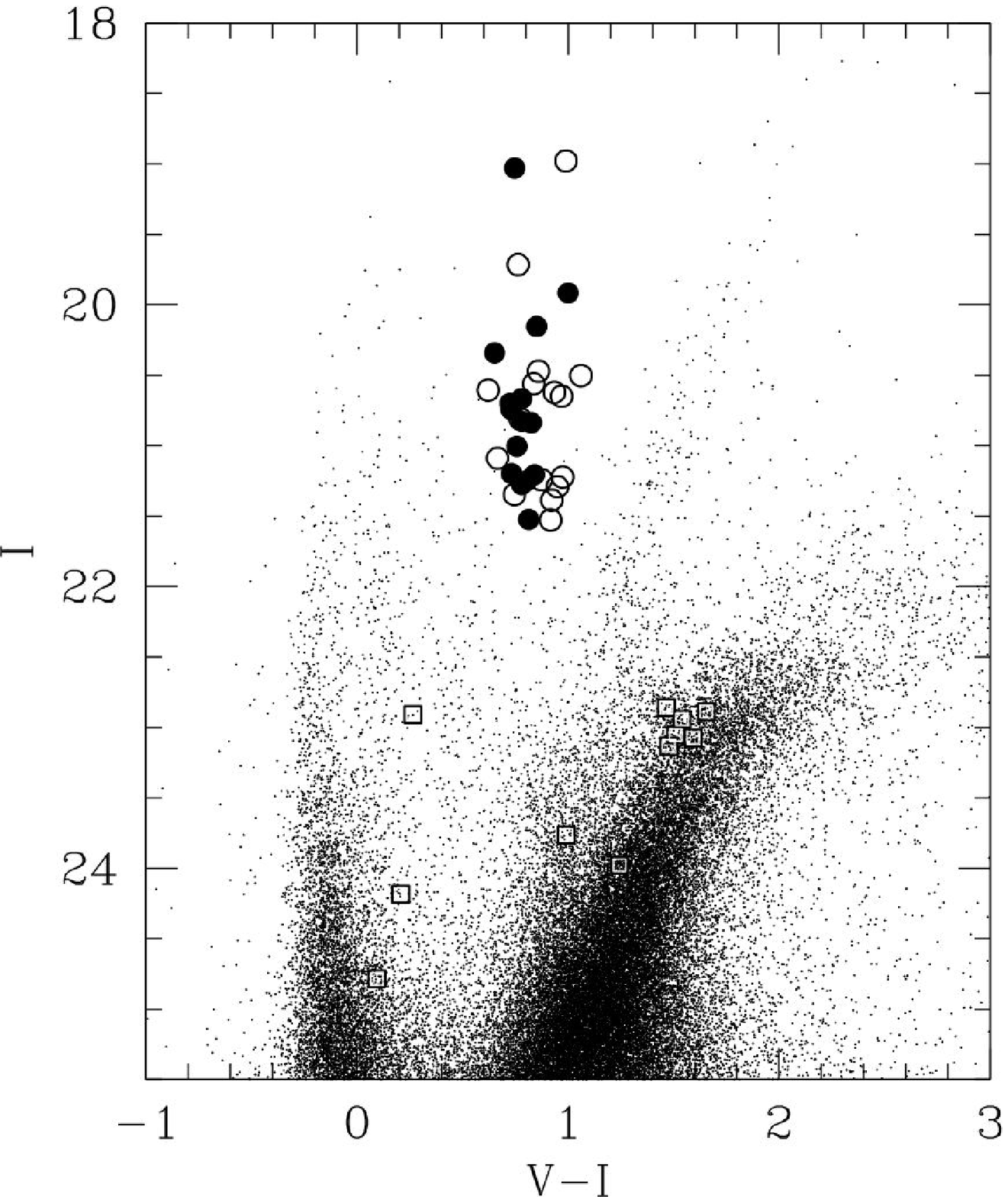} \caption{In this $V$ vs $V-I$ color
magnitude diagram of a representative ACS field in NGC~300 (one of the
two ACS CCD chips in field~5) we overplot the Cepheids ({\em full
circles:} average magnitudes; {\em open circles:} single-epoch ACS
photometry), and their neighbors located within a circle of radius of
0\farcs6 ({\em open squares}), and brighter than 5\% of the mean flux
of the Cepheids.  } \label{cm} \end{figure}

We can derive the typical value of the projected distance of neighbors
from the Cepheids one should use for evaluating their contribution to
the flux measured from the ground, in order to reproduce the results
obtained with the method described above.  We found that in most cases
a distance of 0\farcs6 is the correct choice, corresponding to
approximately half the FWHM of the typical seeing disk of our
images. In addition, we imposed a rather arbitrary lower limit of 5\%
of the mean Cepheid flux for the companions (to be compared with
values between 4\% and 6\% used by the DIRECT Project team). We found
that in order to best reproduce the results obtained from the
difference HST$-$ground at a single epoch, we also had to include only
stars brighter than $V\sim24.4$ and $I\sim23.0$, which suggests that
stars near or below the approximate detection limit of our
ground-based observations do not contribute significantly to the
blending.  Without these restrictions on the fluxes, we would obtain
in general larger corrections for blending than observed from the
direct comparison obtained earlier from the light curves. However, as
we will see in \S5, the final results obtained in this way do not
differ substantially from those obtained from the comparison with the
light curves. In fact, the number of neighbors detected is rather
small, as can be seen in the color-magnitude diagram of Fig.~\ref{cm}.

\begin{deluxetable}{ccc}
\tablewidth{0pt}
\tablecolumns{3}
\tablecaption{Blends: magnitude offsets\label{blends}}
\tablehead{
\colhead{\phantom{aaaa}ID\phantom{aaaaaa}}            &
\colhead{$\Delta V$\tablenotemark{a} }    &
\colhead{$\Delta I$\tablenotemark{a} }    }
\startdata
\vspace{-2mm}
\\
cep040\dotfill & 0.34 & 0.15 \\
cep045\dotfill & 0.15 & 0.18 \\
cep050\dotfill & 0.52 & 0.45 \\
\enddata
\tablenotetext{a}{HST$-$ground-based magnitude}
\end{deluxetable}

\section{Effects of blending on the Period-Luminosity relations}

We can use the results obtained in the previous section to derive the
effects that previously unrecognized blends can have on the optical PL
relations derived for NGC~300 by G04.  We proceed as done in that
paper, by correcting the $V$ and $I$ magnitudes for extinction, using
$A_V = 3.24\,E(B-V)$ and $A_I = 1.96\,E(B-V)$, and assuming
$E(B-V)=0.025$ (\citealt{burstein84}), although the reddening estimate
has been recently increased by the near-IR study of \citet{gieren05}
to $E(B-V)=0.096$. Next, for our scaled-down sample of Cepheids we fit
linear Period-Luminosity relations in $V$, $I$ and the reddening-free
$V-I$ Wesenheit magnitude $W_I$ [$W_I = I - 1.55\,(V-I)$], both before
and after correcting the Cepheid average magnitudes by the amounts
given in Table~\ref{blends}. In our linear regression, we keep the
slopes fixed to the ones determined by the OGLE~II Project in the LMC
(\citealt{udalski00}) and adopted by G04. We are assuming that the
photometric offsets between the HST and ground-based photometry,
measured at a single epoch, can be applied to the average magnitudes
of the Cepheids.

Because of the strictly differential nature of this test, differences
in the intersect values with respect to those given by
G04 for the complete Cepheid sample are not important. For
the un-corrected (blended) case we obtain:

\begin{displaymath}
V_0 = -2.775\,\log P\, +\, 25.116\, (\pm0.053),
\end{displaymath}
\begin{displaymath}
I_0 = -2.977\,\log P\, +\, 24.633\, (\pm0.043),
\end{displaymath}
\begin{displaymath}
W_I = -3.300\,\log P\, +\, 23.897\, (\pm0.054).
\end{displaymath}\\[-5mm]

The corresponding intersects from G04 are 25.155 ($\pm 0.034$), 24.621
($\pm 0.025$) and 23.802 ($\pm 0.028$), respectively.  Once we correct
for the blending, by making the affected Cepheids fainter by the
amounts given in Table~\ref{blends}, we obtain:

\begin{displaymath}
V_0 = -2.775\,\log P\, +\, 25.179\, (\pm0.055),
\end{displaymath}
\begin{displaymath}
I_0 = -2.977\,\log P\, +\, 24.682\, (\pm0.043),
\end{displaymath}
\begin{displaymath}
W_I = -3.300\,\log P\, +\, 23.924\, (\pm0.045).
\end{displaymath}\\[-5mm]

\begin{figure}
\bigskip
\bigskip
\plotone{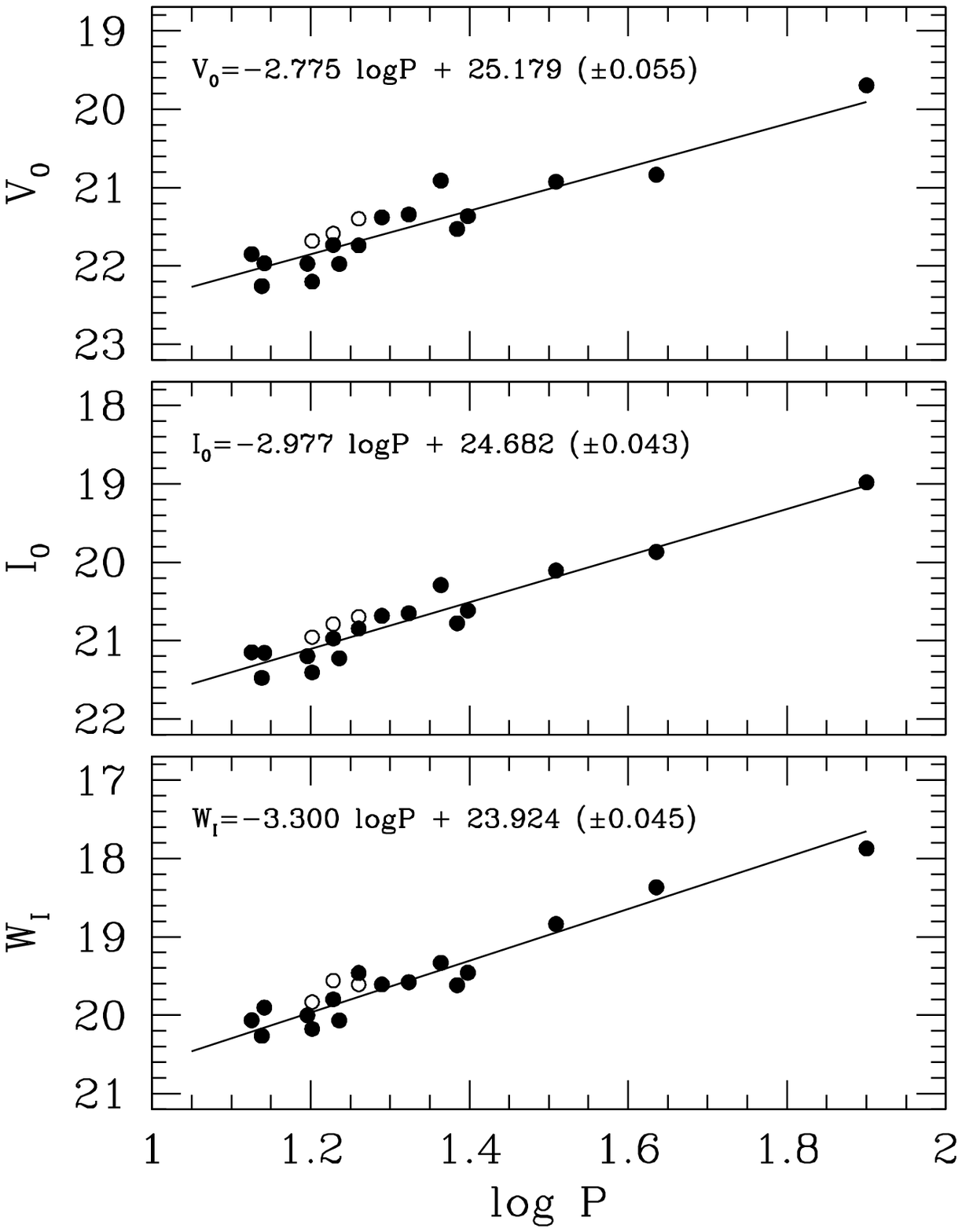}[h]
\caption{Period-Luminosity relations in $V$ ({\em top}), 
$I$ ({\em middle}) and $W_I$ ({\em bottom}) obtained by
a linear regression to the 16 points representing our sample of
Cepheids in NGC~300 for which we have considered the effects of
blending. The open symbols represent magnitudes before the correction
for blending was applied, the full circles represent the same stars
after the correction. The linear fits are done on the corrected sample.
}
\label{fit_corr}
\end{figure}

These regressions  are shown in Fig.~\ref{fit_corr}, where the open
symbols refer to the uncorrected magnitudes, and the full circles
represent the corrected values.


\section{Discussion}

The amounts by which the zero points of the Period-Luminosity
relations increase between the uncorrected case and the case where
blending has been taken into account are 0.063 mag in $V$, 0.049 mag
in $I$ and 0.027 mag in $W_I$. These differences are not statistically
significant, being smaller than or comparable to the size of the errors
in the determination of the intersects. Generalizing to the full
sample of Cepheids available in NGC~300, we conclude that blending
does not appear to have an effect on the distance modulus of NGC~300
which is larger than $\sim 0.04$ magnitudes. Naturally, it would be
desirable to reduce this upper limit even further, but only with the
availability of HST photometry for a considerably larger sample of
Cepheids in NGC~300 we could improve the statistical significance of
our result.

Our result appears quite robust, and does not depend significantly
upon which method is used to estimate the amount of blending. If
instead of the adopted offsets obtained from the single-epoch
comparison with the light curves we sum the flux of all neighbors
projected within 0\farcs6 of the Cepheids, accounting for the 5\%
cut-off, we obtain similar results as above. For instance, the
variation in the zero point between the uncorrected and the corrected
$W_I$ is 0.032 magnitudes, with a 1-$\sigma$ error in the mean of
0.048 mag.

The average amount of blending we found in NGC~300 is smaller than
what the DIRECT Project investigations measured in M31 and M33. For
example, \citet{mochejska00} found a mean $V$-band contribution of
19\% of the Cepheid flux from unresolved neighbors in their sample of
22 variables in M31. An effect of this size would lead to a 9\% upward
correction on the distance to this galaxy. In M33 (64 Cepheids with a
period larger than 10 days) \citet{mochejska01} found a 16\% mean
$V$-band blending effect, leading to a correction of 6\% on the
distance. In NGC~300 the average blending is 7\% (0\% median) for 16
Cepheids, and our estimated upper limit for the effect on the distance
is $\sim$2\%. If we sum the flux of neighbors within 0\farcs6 with a
5\% cutoff (to simulate the Mochejska et al procedure) we obtain an
8\% average blending (6\% median).

The ACS spatial resolution at the distance of NGC~300 (0.5 pc) is
comparable to the resolution of the WFPC2 images used by Mochejska and
collaborators in M31 and M33.  The reason for the different importance
of blending betweeen our study and theirs lies probably in the lower
stellar density in the ACS fields in NGC~300 compared to those
examined in M31 and M33. As shown by \citet{macri01} comparing an
inner NICMOS field in M101 with an outer one, different amounts of
blending are a consequence of varying degrees of crowding and stellar
density.


In G04 we quoted a distance modulus $(m-M)_0 = 26.43 \pm 0.04\,{\rm
(random)} \, \pm\,0.05\,{\rm (systematic)}$. Our estimate for the
systematic error was dominated by the uncertainty on the photometric
zero points, but did not include a contribution from the uncertainty
on the blending. In the discussion presented by G04 we explained why
the effects of blending on our sample of Cepheids in NGC~300 are
likely to be negligible, but we were unable to quantify the size of
the effect or to assign an upper limit to it. Our current work based
on the HST imaging allows us to constrain the maximum effect of
blending, and to add its contribution to the overall error budget.  We
conclude that if blending affects the magnitudes of the Cepheids
included in our Period-Luminosity relation for NGC~300, it does so at
a small level. Its systematic effect on the distance modulus of this
galaxy is limited to less than $\sim0.04$ magnitudes. Our new estimate
for the 1-$\sigma$ systematic uncertainty in the G04 optical distance
modulus to NGC~300 becomes therefore $+0.065$, $-0.05$ magnitudes.

\acknowledgments GP and WG gratefully acknowledge support for this work
from the Chilean FONDAP Center for Astrophysics 15010003 and the
Polish KBN grant No 2P03D02123.

\appendix
\section{ACS photometry of blue supergiants in NGC~300}\label{appendix}

We report in Table~A1 the ACS {\em BVI} photometry of the 40
blue supergiants in common with the catalog by \citet{bresolin02}. The
stellar identification is the same used in that paper.





\begin{deluxetable}{lcrr}
\tablewidth{6cm}
\tablenum{1}
\tablecolumns{4}
\tablecaption{NGC 300 supergiants: ACS photometry\label{phot}}
\tablehead{
\colhead{Star}            &
\colhead{$V$}    &
\colhead{$B-V$}    &
\colhead{$V-I$}    }
\startdata
\vspace{-2mm}
\\
A6  & 19.78 & 0.18    & 0.17    \\
A7  & 20.36 & 0.06    & $-0.03$ \\
A8  & 19.41 & 0.06    & 0.09    \\
A9  & 20.09 & $-0.03$ & $-0.13$ \\
A10 & 18.93 & 0.15    & 0.19    \\
A11 & 18.27 & 0.17    & 0.19    \\
A13 & 19.85 & 0.05    & 0.03    \\
B3  & 20.14 & 0.07    & 0.02    \\
B4  & 20.03 & 0.02    & 0.02    \\
B7  & 20.37 & 0.09    & 0.09    \\
B10 & 19.69 & 0.13    & 0.17    \\
B11 & 19.95 & 0.06    & 0.24    \\
B12 & 19.27 & $-0.04$ & 0.04    \\
B13 & 18.72 & 0.00    & 0.09    \\
B14 & 19.98 & 0.34    & $-0.13$ \\
B15 & 19.71 & 0.01    & 0.01    \\
B17 & 19.53 & 0.02    & 0.06    \\
B18 & 20.25 & $-0.06$ & $-0.07$ \\
B19 & 20.23 & 0.07    & 0.21    \\
C1  & 18.99 & 0.18    & 0.12    \\
C5  & 20.77 & $-0.01$ & $-0.08$ \\
C6  & 19.84 & 0.12    & 0.15    \\
C8  & 20.01 & 0.06    & 0.05    \\
C9  & 20.22 & 0.09    & 0.09    \\
C11 & 19.79 & $-0.01$ & 0.02    \\
C12 & 20.34 & 0.01    & 0.27    \\
C14 & 20.25 & $-0.03$ & $-0.12$ \\
C15 & 21.03 & $-0.06$ & $-0.15$ \\
C17 & 20.86 & 0.27    & 0.26    \\
C18 & 19.94 & $-0.13$ & $-0.17$ \\
C19\footnote{This star was incorrectly assigned the photometry of a fainter,
nearby star by \citet{bresolin02}.} & 19.76 & $-0.02$ & $-0.10$ \\
D1  & 20.94 & 0.17    & 0.13    \\
D2  & 19.83 & 0.13    & 0.19    \\
D3  & 21.71 & 0.17    & 0.15    \\
D5  & 20.06 & 0.09    & 0.08    \\
D12 & 18.66 & 0.31    & 0.31    \\
D13 & 18.86 & 0.10    & 0.14    \\
D16 & 20.84 & 0.20    & 0.20    \\
D17 & 19.44 & 0.15    & 0.11    \\
D18 & 19.96 & 0.05    & 0.07    \\
\enddata
\end{deluxetable}

\clearpage


\end{document}